\pdfoutput=1 
\documentclass[aps,prl,twocolumn]{revtex4-1}
\usepackage{blindtext}
\usepackage{graphicx}
\usepackage{amsmath}
\usepackage{multirow}
\usepackage{hyperref}
\usepackage{amssymb}
\usepackage{color}
\usepackage{braket}

\newcommand{\rvec}{{{\bf r}}}
\newcommand{\kvec}{{{\bf k}}}
\newcommand{\Pvec}{{{\bf P}}}
\newcommand{\Svec}{{{\bf S}}}
\newcommand{\kappavec}{{{\boldsymbol \kappa}}}
\newcommand{\etaSOC}{\eta_{soc}}

\begin{document}

\title{Finite Temperature Phases  of Two Dimensional Spin-Orbit Coupled Bosons}
\author{Eiji Kawasaki$^1$}
\email{eiji.kawasaki@grenoble.cnrs.fr}
\author{Markus Holzmann$^{1,2}$}
\email{markus.holzmann@grenoble.cnrs.fr}
\affiliation{$^1$LPMMC, UMR 5493 of CNRS, Universit{\'e} Grenoble Alpes, BP 166, 38042 Grenoble, France}
\affiliation{$^2$Institut Laue-Langevin, BP 156, F-38042 Grenoble Cedex 9, France}

\begin{abstract}
We determine the finite temperature phase diagram of two dimensional bosons with two hyperfine (pseudo-spin) states coupled via Rashba-Dresselhaus spin-orbit interaction using classical field Monte Carlo calculations. 
For anisotropic spin-orbit coupling, we find a transition to a Berenzinskii-Kosterlitz-Thouless superfluid phase with quasi-long range order. We show that the spin-order of the quasi-condensate is driven by the anisotropy of interparticle interaction, favoring either a homogeneous plane wave state or stripe phase with broken translational symmetry. Both phases show characteristic behavior in the algebraically decaying spin density correlation function. For fully isotropic interparticle interaction, our calculations indicate a fractionalized quasi-condensate where the mean-field degeneracy of plane wave and stripe phase remains robust against critical fluctuations. In the case of fully isotropic spin-orbit coupling,  the circular degeneracy of the single particle ground state destroys the algebraic ordered phase in the thermodynamic limit, but a cross-over remains for finite size systems.
\end{abstract}

\maketitle
	\textbf{Introduction.} The coupling of artificial gauge fields to ultracold atomic gases \cite{RevModPhys.83.1523} has opened the possibility of studying spin-orbit coupled Bose gases \cite{Lin2011,PhysRevLett.114.125301,Galitski2013,revew}
where translational symmetry may be broken spontaneously in the superfluid ground state \cite{2016arXiv161008194L}. At the mean-field level, spin-orbit coupling (SOC) introduces degenerate ground states   expected to enhance fluctuation effects and giving rise to new, exotic quantum  phases. 
The occurrence and nature of finite temperature transitions in these systems have not yet been fully established 
\cite{2016arXiv160905464S,PhysRevLett.109.025301,PhysRevLett.110.085304,0034-4885-78-2-026001,PhysRevB.84.060508,
PhysRevA.89.063614}.  
	
	In the following we consider a two-dimensional homogeneous gas of Rashba-Dresselhaus spin-orbit coupled bosons. Mean-field calculations \cite{0034-4885-78-2-026001,PhysRevB.84.060508,PhysRevA.89.063614,PhysRevLett.107.150403,PhysRevLett.105.160403} indicate a Bose condensed ground state of a single plane wave with non-vanishing momentum or a linear superposition of two plane waves with opposite momenta, called plane wave state (PW) and stripe phase (SP), respectively. For spin-independent particle interaction, PW and SP remain degenerate at the mean-field level. In addition, in the case of isotropic SOC, ground states with momenta lying on a circle are connected by symmetry. These degeneracies may be broken by classical or quantum fluctuations.
	
In this work we explore the phase diagram using classical field Monte Carlo calculations. We show that for anisotropic SOC, the systems undergoes a Kosterlitz-Thouless phase transition from a normal to superfluid state.
In the superfluid state, the single particle density matrix decays algebraically and directly reflects the PW/SP character of the mean-field ground state. In the limit of isotropic interparticle interaction, the PW/SP degeneracy is unaffected by the transition. Thus, at large but finite system sizes, fragmentation \cite{PhysRevA.74.033612} of the condensate  occurs.
In the case of isotropic SOC, we show that the transition temperature decreases with increasing system size due to the increasing number of degenerate mean-field ground states and eventually vanishes in the thermodynamic limit.

Since continuous phase transitions at finite temperatures are driven by classical long wave length fluctuations, our classical field calculations allows us to establish quantitatively the finite temperature phase diagram in a parameter regime relevant for ultracold atom experiments.

	\textbf{Model.} We consider bosons in two hyperfine states, $\sigma$, labeled $\uparrow$ and $\downarrow$, 
	described by the Hamiltonian $H=H_0+V$, where $H_0$ describes ideal spin-orbit coupled bosons and $V$ the interparticle interaction. We have
	\begin{equation}
	H_0=\int d^2\rvec \, \hat{\Psi}^\dagger(\rvec) \left[ -\frac{\hbar^2\nabla^2}{2m} - i \frac{\hbar^2 \kappa}{m} \left(\sigma_x \partial_x + \etaSOC \sigma_y \partial_y \right) \right] \hat{\Psi}(\rvec)
    \end{equation}	
	where $\hat{\Psi}^\dagger \equiv (\hat{\Psi}_\uparrow^\dagger , \hat{\Psi}_\downarrow^\dagger )$ is the field creation operator, $\sigma_\alpha$ are the Pauli matrices, and $m$ is the atomic mass. The SOC is characterized by its strength $\kappa$ and the anisotropy $0\le \etaSOC \le 1$. For anisotropic SOC, $\etaSOC <1$, the minimum of the single particle energies is reached at two wave vectors $(\pm \kappa,0)$, whereas all wave vectors on a circle of absolute value $\kappa$ are degenerate for isotropic SOC $\etaSOC=1$.
	
	The interparticle interaction is described by
\begin{equation}
    V=\frac12 \sum_{\sigma,\sigma'=\uparrow \downarrow} g_{\sigma \sigma'}
    \hat{\Psi}^\dagger_\sigma(\rvec)  \hat{\Psi}^\dagger_{\sigma'}(\rvec) \hat{\Psi}_{\sigma'}(\rvec) \hat{\Psi}_{\sigma}(\rvec)
\end{equation}	
where the coupling strengths $g_{\sigma \sigma'}$ are determined by the scattering amplitudes of different hyperfine states. For simplification, we only consider $g_{\uparrow \uparrow}=g_{\downarrow \downarrow}$ in the following and use $g=g_{\uparrow \downarrow}-g_{\uparrow \uparrow}$ to characterize the interaction imbalance.
	
The mean field ground state wave function can be written in terms of
\begin{eqnarray}
\psi^{MF}_{\kappavec} = \frac1{\sqrt{2}} \left[ \cos(\phi) e^{i \kappavec \cdot \rvec}
		\left(
		    \begin{array}{c}
		       1 \\
		      -e^{i \theta_\kappavec}
		    \end{array}
		  \right)
 + \sin(\phi) e^{-i \kappavec \cdot \rvec}
 		\left(
		    \begin{array}{c}
		     1 \\
		      e^{i \theta_\kappavec}
		    \end{array}
		  \right) \right]
		  \nonumber \\
\end{eqnarray} 	
with $|\kappavec|=\kappa$ and $e^{i \theta_\kvec}\equiv (k_x+ i \etaSOC k_y)/\sqrt{k_x^2+ \etaSOC^2 k_y^2}$. The circular degeneracy of the wave vector is broken for anisotropic SOC where $\kappavec$ points along the $x$ direction.  For $g<0$, the mean field energy is minimized by a single plane wave state, $\phi=0$ (PW), whereas stripes corresponding to $\phi=\pi/4$ (SP) are formed in the real space density for $g>0$. The mean-field energy is independent of $\phi$ for isotropic interactions, $g=0$; PW and SP are degenerate at the mean-field level in this case.

Expanding the effective action at low temperatures around the mean-field state at $g \ne 0$ to include thermal fluctuations \cite{PhysRevB.84.060508,PhysRevA.89.063614}, a Kosterlitz-Thouless transition is expected for anisotropic SOC, but the occurrence of a superfluid phase for $\etaSOC \to 1$ is controversial. Further, in the limit of isotropic interaction, $g=0$, the effective action approach breaks down, since PW and SP become degenerate at the mean-field level and the character of the low temperature phase is still an open problem. To overcome these limitations, we have performed classical field Monte Carlo calculations which are expected to correctly describe the finite temperature behavior close to a possible phase transition where quantum fluctuations should not play an essential role and can therefore be neglected \cite{PhysRevLett.83.1703,PhysRevLett.90.040402,Svistunov,PNAS}.

We have calculated the reduced single particle density matrix, $G_{\sigma, \sigma'}(\rvec,\rvec')= \langle \hat{\Psi}_{\sigma'}^\dagger(\rvec')\hat{\Psi}_\sigma(\rvec)\rangle$ expected to reveal algebraic order for a quasi-Bose condensed state \cite{Berenzinskii}.
Superfluid mass density, $\rho_s$, can be directly related to the phase stiffness,
$\rho_s=\partial^ 2 F(\theta)/\partial \theta^2$, where $F(\theta)$ is the free energy density where the momentum operator, $\hat{p}$, is replaced by $\hat{p} - \theta$ in the Hamiltonian \cite{Baym69,PhysRevB.36.8343,PRB}.

The spin-order of the low temperature phase is characterized by the spin-density correlation functions, $M_\alpha(\rvec)=\langle \hat{\Svec}_\alpha(\rvec) \hat{\Svec}_\alpha(0)\rangle$,  where $\hat{\Svec}_\alpha(\rvec)=\hat{\Psi}^\dagger(\rvec) \sigma_\alpha \hat{\Psi}(\rvec)$ is the local magnetization operator. From the mean-field solution, we expect $M_{\alpha}(\rvec)$ to monotonically reach a constant value  at large distances for PW states when $\alpha$ points in the direction of $\kappavec$. In the stripe phase oscillating behavior occurs, since spins rotate in the plane orthogonal to the vector $\kappavec$ with a spatial frequency equal to $2\kappa$~\cite{PhysRevA.85.063623}. 
 
\textbf{Method.} In the following we are interested to establish the phase diagram in the limit of small interaction strength, $m g \ll 1$ and small spin-orbit coupling, $\kappa \lambda_T \ll 1$ where $\lambda_T=\sqrt{2 \pi \hbar^2/m k_B T}$ is the thermal wave length at temperature $T$. In this limit, the leading order corrections to mean-field are captured within classical field theory \cite{PhysRevLett.83.1703,PhysRevLett.90.040402,PhysRevA.76.013613,Svistunov,PNAS} where the occupation of low energy modes is high such that commutators like $[\hat{\Psi}^\dagger(\rvec),\hat{\Psi}(\rvec')]$ can be neglected. In this approximation, the field operator, $\hat{\Psi}(\rvec)$, can be replaced by two complex fields, $\Psi(\rvec)\equiv (\Psi_\uparrow(\rvec),\Psi_\downarrow(\rvec))$, one for each spin, and the theory is regularized by discretizing space on a lattice of linear extension $L$.

The probability of a given field configuration is then proportional to $\exp(-S[\Psi(\rvec)])$ and the action writes
	\begin{eqnarray}
		 S[\Psi(\rvec)]  &=& \frac{a^2}{k_B T} \sum_{\rvec} \Big\{
		 \sum_{\sigma=\uparrow,\downarrow}\left[-\Psi_{\sigma}^*(\rvec)\frac{\hbar^ 2\nabla^2_D}{2m}\Psi_{\sigma}(\rvec)-\mu|\Psi_{\sigma}(\rvec)|^2\right] \nonumber \\
		 		&&+ \frac{\hbar^2\kappa}{m} \left[\Psi_{\uparrow}^*(\rvec)\left(-i\partial_{x}^{D}-\etaSOC\partial_{y}^{D}\right)\Psi_{\downarrow}(\rvec)\right] \nonumber \\
		&&+ \frac{\hbar^2 \kappa}{m} \left[\Psi_{\downarrow}^*(\rvec)\left(-i\partial_{x}^{D}+\etaSOC\partial_{y}^{D}\right)\Psi_{\uparrow}(\rvec)\right] \nonumber\\
		&&+ \frac12 \sum_{\sigma,\sigma'=\uparrow,\downarrow} g_{\sigma\sigma'}|\Psi_{\sigma}(\rvec)|^2|\Psi_{\sigma'}(\rvec)|^2 \Big\}
		\label{local_action}
	\end{eqnarray}
where $a$ is the lattice spacing, $\mu$ is the chemical potential, and $\nabla_D$ and $\partial_{\alpha}^{D}$ are finite difference expressions approximating the derivatives. It can be shown that the action is real for all field configurations so that the partition function can be sampled by Monte Carlo methods. We have implemented local moves in real and Fourier space to efficiently sample around the mean-field energy minima.

Within classical field theory, the occupation of eigenmodes of energy $\varepsilon$ is by the equipartition theorem, $n_{cf}(\varepsilon)=k_B T/\varepsilon$ instead of the full Bose distribution, $n_B(\varepsilon)=[\exp(\varepsilon/k_B T)-1]^{-1}$. Since we expect mean-field theory to quantitatively describe high energy modes, we have corrected the densities of our classical field calculations to account for the correct ultraviolet behavior \cite{PNAS} adding the difference 
\begin{equation}
\Delta n=
\frac1{L^2} \sum_{\kvec,\alpha=\pm } \left[
 n_B(\varepsilon_{\kvec \alpha}^{mf}-\mu) - n_{cf}(\varepsilon_{\kvec \alpha}^{mf}-\mu)
  \right]
\end{equation}
Here, the single particle mean-field energies are given by $\varepsilon_{\kvec \alpha}^{mf}= \varepsilon_{\kvec \alpha} + 2 \sum_{\alpha'} g_{\alpha \alpha'} n_{\alpha'}^{mf}$, where $\varepsilon_{\kvec \alpha}$ are the  eigen energies of the ideal SOC gas. The corresponding mean-field densities, $n_\alpha^{mf} = L^{-2} \sum_{\kvec} n_{cf/B}(\varepsilon_{\kvec \alpha}^{mf}-\mu)$, have to be determined self-consistently
 
 \textbf{Results.} In order to study the competition between SOC and interparticle interaction, we have fixed $\sum_{\sigma \sigma'=\uparrow \downarrow} m g_{\sigma \sigma'}/4 = \kappa/\sqrt{m k_B T}= \pi/20$ with $mg=0$ to address isotropic  interaction and $mg = \pm \pi/100$ to slightly break the spin isotropy of scattering particles. In the following we will study the phase diagram as a function of the SOC anisotropy $\etaSOC$ and observe signatures of the different phases in the condensate and superfluid fraction and in the spin-resolved density correlation function.
 
In two dimensions, without SOC,  the single particle density matrix, $G_{\sigma, \sigma'}(\rvec,\rvec')$, is expected to decay algebraically in the low temperature superfluid phase \cite{Berenzinskii,KT,Kosterlitz}. For SOC bosons, quasi-long range order for SOC bosons occurs when the single particle density matrix is dominated by one or few highly occupied modes. We therefore project $G
_{\sigma, \sigma'}(\rvec,\rvec')$ over all degenerate PW mean-field ground states
\begin{equation}
n_0^\kappa= \sum_{\kvec=(\pm \kappa,0)} \sum_{\sigma \sigma'} \int \frac{d\rvec d\rvec'}{L^2} \psi^{MF}_{\kvec  \sigma}(\rvec) G_{\sigma, \sigma'}(\rvec,\rvec') \psi^{MF*}_{\kvec  \sigma'}(\rvec') 
\end{equation}
to estimate the condensate fraction $n_0^\kappa/n$ where $n=\sum_\sigma G_{\sigma \sigma}(\rvec,\rvec)$ is the total particle density. Although $n_0^\kappa$ is a direct indicator for a phase transition it does not distinguish PW or SP character. 

	\begin{figure}[ht]
		\includegraphics[scale=.65]{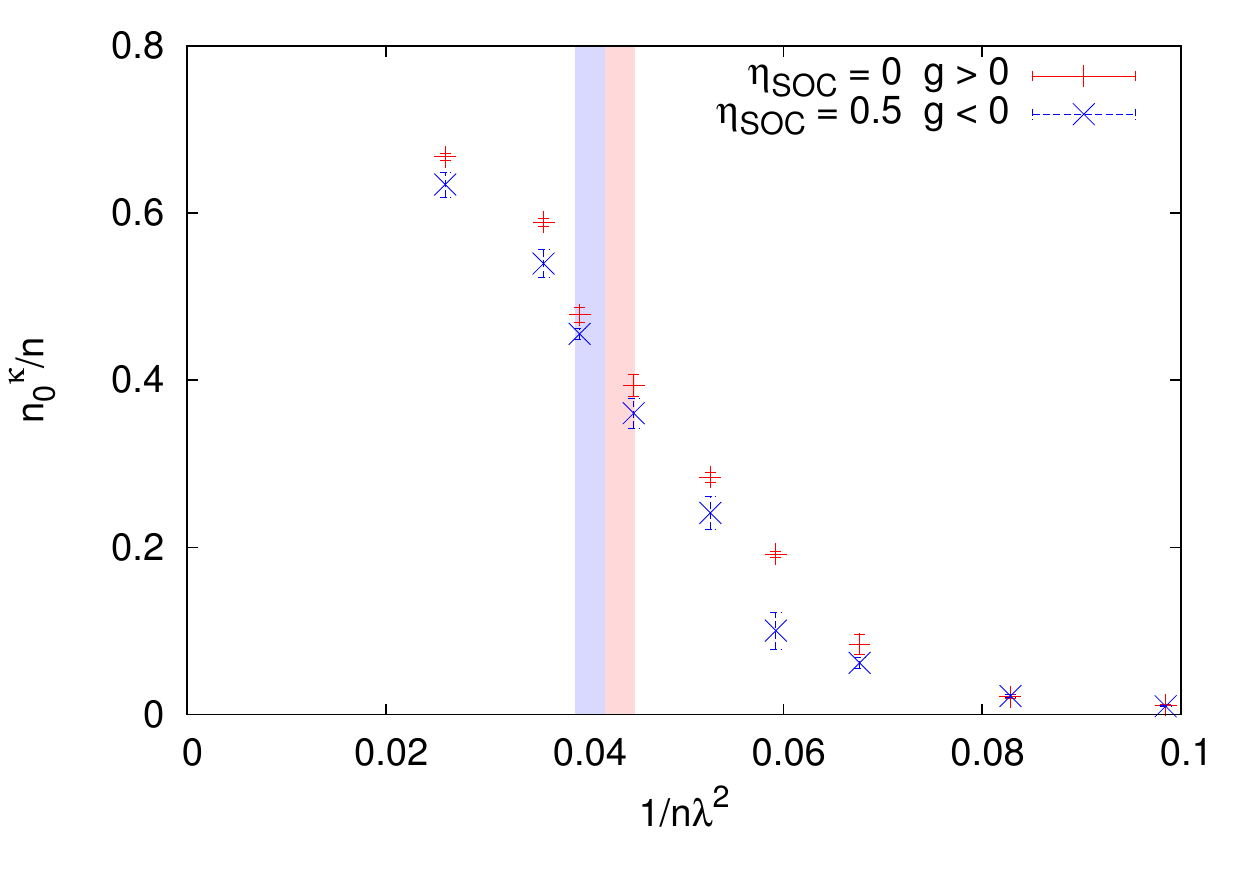}
		\caption{Condensate fraction $n_0^\kappa/n$ as a function of inverse phase space density $[n \lambda^2]^{-1}$ for a finite system of length $L/a=80$. The cross-over from normal to condensed phase slightly lowers with increasing SOC anisotropy, $\etaSOC$. Although the PW/SOC character of the condensate depends essentially on the sign of the anisotropic interaction $m g=\pm \pi/100$, differences in $n_0^\kappa$ between $g\ge 0$ and $g<0$ for equal SOC  are beyond our resolution. The colored zones indicate our estimates for the Kosterlitz-Thouless transition in the thermodynamic limit.}
		\label{BKT_bis}
	\end{figure}

In Fig.~\ref{BKT_bis} we show $n_0^\kappa$ as a function of density for a finite system of extension $L/a=80$ with $a=\hbar/\sqrt{m k_B T}$.
The condensate fraction grows rapidly around a cross-over density which decreases  from $\etaSOC=0$ to $\etaSOC=0.9$ \footnote{Notice that for $\etaSOC=0$ the SOC can be eliminated via spin-rotation combined with a gauge transformation whenever $\kappavec$ is commensurate with the boundary conditions. The resulting $\psi^4$ theory with $N=4$ field components does not show a Kosterlitz-Thouless phase transition. Here, we address the limit $\etaSOC \to 0$ continuously connected to non vanishing $\etaSOC >0$, which corresponds to non commensurate values of $\kappavec$.}. However, no differences are visible changing the sign of our small anisotropic interaction from negative to positive $g$. We therefore expect that the cross-over/transition temperature is a smooth, continuous function of $g$ around $g=0$.

	\begin{figure}
		\includegraphics[scale=.65]{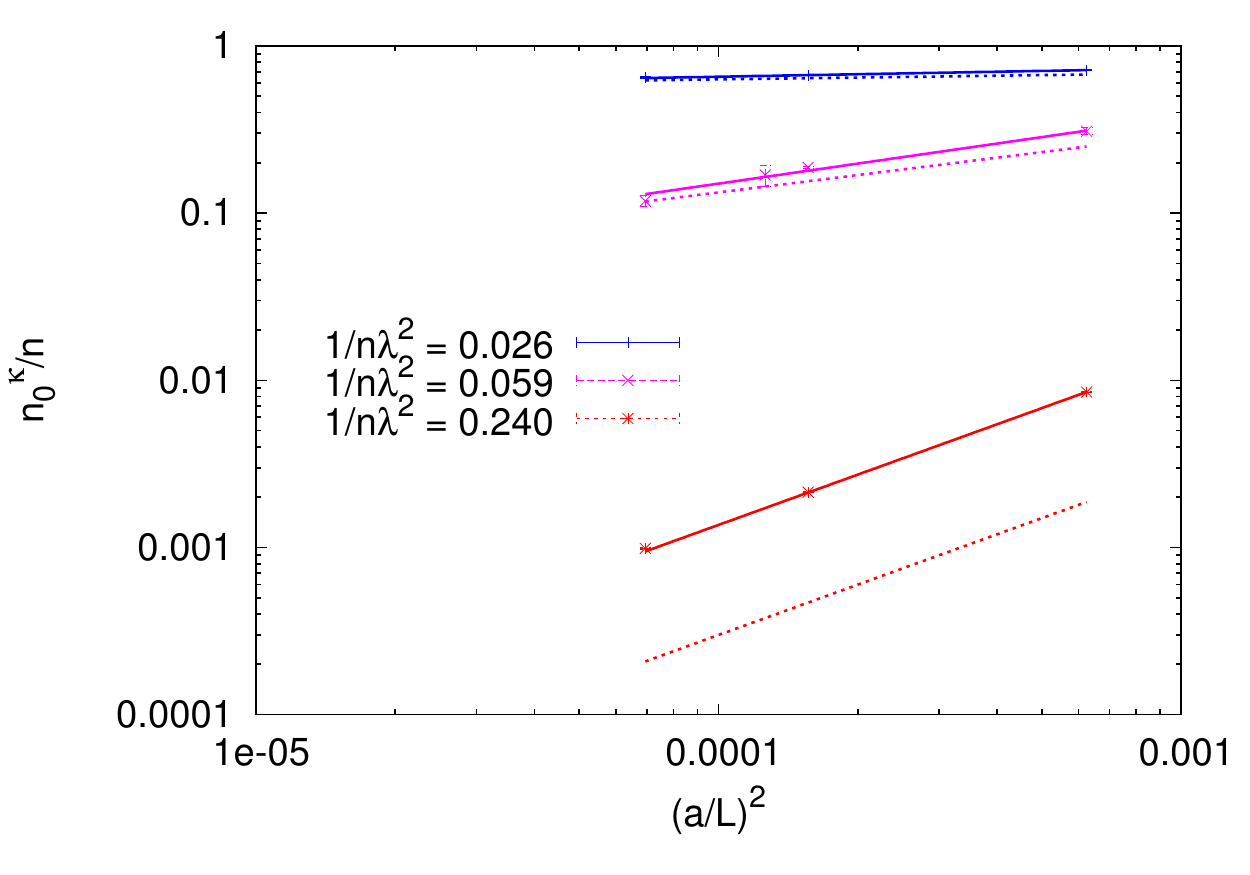}
		\caption{Solid lines: Condensate fraction, $n_0^\kappa/n$, as a function of the inverse volume, $L^{-2}$, for anisotropic SOC bosons with $\etaSOC=0$ at different phase space densities and anisotropic interaction, $g >0$. Dashed lines show the corresponding maximal occupation number after diagonalizing the single body density matrix (not ensemble averaged). In the normal phase at low phase space density, we have $n_0^\kappa \sim L^{-2}$ and two degenerate modes, whereas in the superfluid phase at high phase space density we have $n_0^\kappa \sim L^{-(2-\eta)}$ with $\eta >1/4$, the degeneracy is broken, and only one mode contributes to the quasi-condensate.}
		\label{BKT}
	\end{figure}
	
 In order to determine a possible sharp phase transition in the thermodynamic limit, we have determined the behavior of the condensate fraction increasing the system size, $n_0^\kappa/n \sim L^{-(2-\eta(T))}$. Whereas in the high temperature, normal phase the condensate fraction decreases with the volume, $\eta(T)=0$, the exponent changes in the low temperature phase indicating a Berenzinskii-Kosterlitz-Thouless transition \cite{Berenzinskii,KT,Kosterlitz}. Assuming the transition to be within the Kosterlitz-Thouless class, the critical temperature can be estimated to occur when $\eta(T_{KT})=1/4$. Our calculations indicate that for anisotropic SOC, $\etaSOC<1$ the Berenzinskii-Kosterlitz-Thouless phase occurs at finite temperature, independent of the sign of $g$ (see Fig.~\ref{BKT} for $\etaSOC=0$). Further, the limit of isotropic interaction, $g=0$, is indeed approached smoothly from both sides, $g>0$ and $g<0$.
 
\begin{figure}[ht]
		\includegraphics[scale=.65]{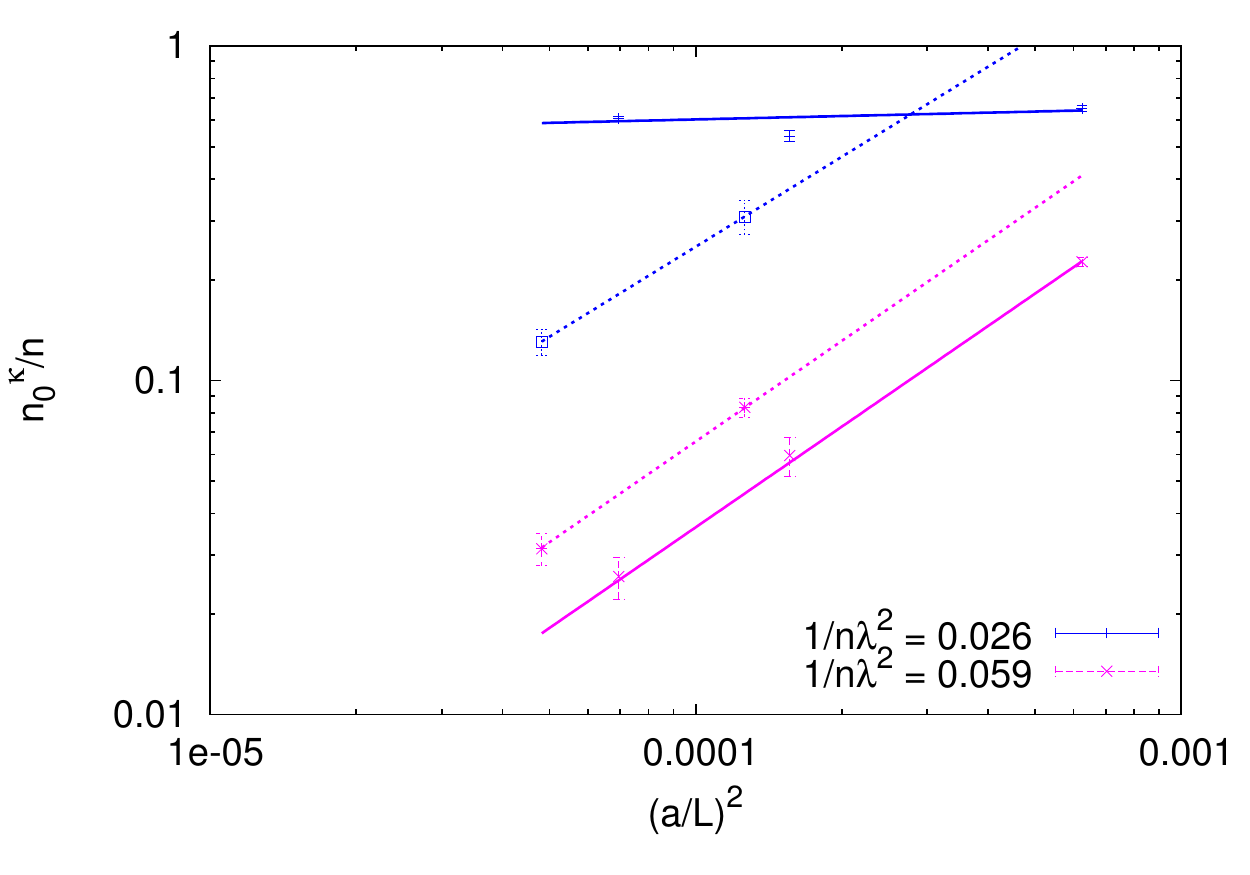}
		\caption{Solid lines: Condensate fraction, $n_0^\kappa/n$, as a function of inverse volume for isotropic SOC, $\etaSOC=1$, of finite systems with 4 degenerate minima and $g>0$. Dashed lines corresponds to finite systems with 8 degenerate minima where the algebraic behavior, $n_0^\kappa \sim L^{-(2-\eta)}$ with $\eta>1/4$, at high phase space density is suppressed.
		}
		\label{eta1_BKT}
	\end{figure}

For isotropic SOC, $\etaSOC=1$, however, we do not observe the onset of quasi-long range order for $g<0$ in the considered density regime and system sizes. For $g>0$, a cross-over occurs, but 
the onset of algebraic order strongly depends on the number of degenerate mean-field ground states. Although these energy minima form a circle in the thermodynamic limit, in a finite geometry only a certain number of single particle states are strictly degenerate. As shown in Fig.~\ref{eta1_BKT}, the behavior of the condensate fraction is qualitatively and quantitatively affected by the number of degenerate states. 
 In particular, the onset of algebraic order is shifted towards considerable higher densities (lower temperatures) increasing the degeneracy from $4$ to $8$ degenerate modes.  
For $\etaSOC=1$ and infinity system sizes, the transition will therefore be shifted to zero temperature and no finite temperature transition with algebraic order in the single particle channel should occur due to the circular degeneracy.

We have further calculated the superfluid and  normal  mass density, $\rho_n=m n-\rho_s$ from the phase stiffness
\begin{equation}
\rho_n= \frac1{k_B TL^2} \langle\left[\Pvec^{\text{tot}}_\alpha + \hbar \kappa \Svec^{\text{tot}}_\alpha \right]^2 \rangle
\end{equation}
where $\Pvec_\alpha^\text{tot}$ is the total momentum and $\Svec_\alpha^\text{tot}= \int d\rvec S_\alpha(\rvec) $ the total magnetization of the system in the $\alpha=x/y$ direction. 
Deviations from a Boltzmann distribution of $[ \Pvec^{\text{tot}}_\alpha + \hbar \kappa \Svec^{\text{tot}}_\alpha ]^2/(2m n L^2)$ are directly connected to the quantization of the center of mass motion.
Consistent with the prediction of Berenzinskii, Kosterlitz and Thouless, the low temperature, algebraically ordered phase is superfluid for $\etaSOC<1$ whereas it vanishes for isotropic SOC with increasing degeneracy. Josephson's scaling relates the algebraic decay of the single particle density matrix to the condensate and superfluid density \cite{Josephson,PRB}, and our values of $\eta(T)$ from $n_0^\kappa$ are consistent with  $\eta(T)=m/\rho_s\lambda^2$ within the numerical uncertainty.

The absence of a transition for isotropic SOC is consistent with rather general considerations particular to two dimensional systems.
From the analysis of the non-linear $\sigma$-model \cite{Polyakov,PhysRevB.14.3110}, a phase transition of the Kosterlitz-Thouless type is only expected for $N=2$ component fields,
e.g. real and imaginary part of a spinless complex field.
For $\etaSOC=1$, the PW (SP) order parameter of the infinite system is characterized
its phase and 
by its spin-direction, $-\kappavec/\kappa$, and can be mapped to the two-dimensional (half) unit-sphere.
Therefore, vortices are not any more topologically stable objects and destroy algebraic order
and superfluidity at any finite temperature.
For $\etaSOC < 1$, the two possible spin-directions are disconnected, so that algebraic order and  superfluidity is protected against vortex excitations at low temperatures.


We now want to characterize the order -- PW or SP -- of the superfluid phase for $\etaSOC <1$. Therefore, we diagonalize the single particle density matrix, $G_{\sigma \sigma'}(\rvec,\rvec')$, calculated by a single realization, not ensemble averaged over different initial conditions. The spin structure of the eigenmodes of $G$ is resolved by taking into account the degenerate Fourier modes with $\kvec=(\pm \kappa,0)$. As shown in Fig~\ref{BKT}, for $g \ne 0$, above the critical density, the single particle density matrix is dominated by a single, highly occupied mode, yielding PW and  SP order for $g<0$ and $g>0$, respectively.

\begin{figure}[ht]
		\includegraphics[scale=.65]{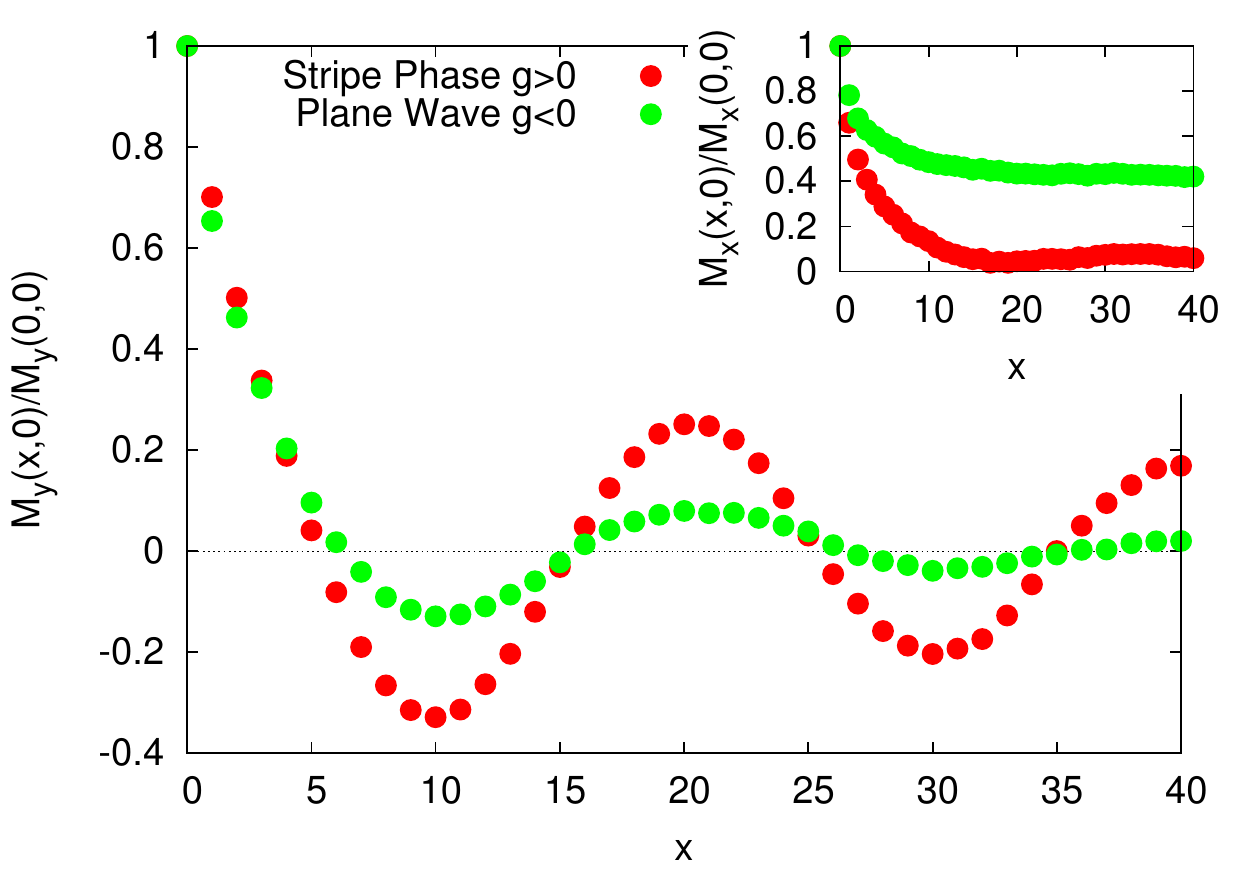}
		\caption{Spins density correlation function, $M_\alpha(x,0)$, at phase space density $1/n\lambda^ 2\simeq 0.026$ for $\etaSOC=0.9$ where $n_0^\kappa/n\sim 40\%$. For $g<0$, $M_x(x,0)$ shows quasi-long range order indicating PW, whereas $M_y(x,0)$ is short ranged. For $g>0$ we obtain SP where the amplitude of the oscillations of $M_y(x,0)$ decay algebraically and no order is present in $M_x(x,0)$. }
		\label{SPcorr}
	\end{figure}

The spin structure of this condensate mode is directly reflected in the spin correlation function, $M_\alpha(\rvec)$, shown in Fig.~\ref{SPcorr}. For $g<0$, $M_x(x,0)$ develops quasi-long range order, whereas it remains short ranged in the SP for $g>0$. Quasi-long range stripe order is reflected in slowly decaying oscillations of period $2 \kappa$ in $M_y(x,0)$. In both cases, the exponent of the algebraic decay is given by the scaling exponent of $n_0^\kappa$ and  compatible with $\eta(T)$ obtained from the superfluid density. Therefore, the quasi-long range spin order results from the spin structure of the underlying quasi-condensate. 

In the limit of isotropic interaction, $g=0$, we always obtained two highly occupied modes of the single particle density matrix, degenerate within our numerical precision. Therefore, PW and SP remain degenerate and robust against thermal, critical fluctuations and we obtain a fractionalized quasi-condensate. Both spin correlation functions, $M_x(x,0)$ and $M_y(x,0)$ become quasi-long ranged and indicate simultaneous PW and SP character.
From the Bogoliubov approximation around the $T=0$ mean-field ground states, we expect that quantum fluctuations lift the degeneracy and favor the PW character decreasing the temperature without further phase transition \cite{PhysRevA.85.023615}. 

	\textbf{Conclusion.}
	Using classical field simulations we have studied the phase diagram of two-dimensional, spin-orbit coupled Bosons as a function of the spin-orbit anisotropy $\etaSOC$ with vanishing or small spin-anisotropy, $g$, in the interparticle interaction. For $\etaSOC<1$, our calculations are consistent with a Berenzinskii-Kosterlitz-Thouless transition. The low temperature superfluid phase is characterized by the PW or SP character of the underlying quasi-condensate determined by the sign of $g$. For isotropic interactions, $g=0$, we obtained a fractionalized quasi-condensate with two degenerate modes at the transition showing both, PW and SP character. For fully isotropic SOC, $\etaSOC=1$, a cross-over occurs for finite systems, but no superfluid transition is expected in the thermodynamic limit.

	\textbf{Acknowledgments.} We thank Gordon Baym for seminal discussions, Frank Hekking and Tomoki Ozawa for useful comments and a critical reading of the manuscript,  and Francesco Calcavecchia for continuous support and discussions.

\bibliography{article}

\end{document}